# An Energy-Efficient and Runtime-Reconfigurable FPGA-Based Accelerator for Robotic Localization Systems

Qiang Liu*[1], Zishen Wan*[2], Bo Yu*[3], Weizhuang Liu[1], Shaoshan Liu[3], Arijit Raychowdhury[2]

[1]Tianjin University, China, [2]Georgia Institute of Technology, USA, [3]PerceptIn, USA

*Equally-Credited Authors (ECAs)

A robot usually localizes itself in an environment by estimating the collection of its position and rotation states, while constructing a map of unknown surroundings, giving rise to the notion of Simultaneous Localization and Mapping (SLAM). SLAM is a fundamental kernel in autonomous machines at all computing scales, from drones, AR, VR to self-driving cars. Principled mathematical solutions for SLAM involve filtering-based or non-linear optimization-based (Fig. 1a), where the latter recently shows higher robustness but with intensive computation. Prior ASICs [1,2] and FPGAs [3,4,5] have accelerated SLAM on hardware, but they usually target one specific design. In this work, we present a runtime-reconfigurable FPGA accelerator for robotic localization tasks. We exploit SLAM-specific data locality, sparsity, reuse, and parallelism, and achieve >5x performance improvement over the state-of-the-art. Especially, our design is reconfigurable at runtime according to the environment and platform to save power while sustaining accuracy and performance.

Fig.1b shows the SLAM system compute latency characterization on software. SLAM consists of a vision frontend to extract features and an optimization backend to estimate the pose. We find that the localization computation accounts for 46% and 78% latency on two commonly-used SLAM systems, indicating a lucrative acceleration target. The localization is usually formulated as a constrained non-linear optimization problem, often through bundle adjustment, which minimizes the pose projection errors from 2D features to 3D points in the map. The optimization problem is solved using the Levenberg-Marquardt (LM) method, consisting of 1) a nonlinear least squares (NLS) solver that solves maximum a posteriori estimation, and 2) marginalization that generates the prior of NLS solver. We will accelerate both phases through software-hardware co-design by leveraging SLAM-specific data patterns and inherent parallelism.

The proposed robotic localization design accelerates both NLS solver and marginalization algorithm (Fig. 2a). The NLS solver circuit first calculates Jacobians, following by Schur elimination and Cholesky decomposition. Marginalization is performed after NLS solver. Fig. 2b shows the circuit for visual Jacobian. We divide the computation into three levels: keyframe, feature, and observation. The keyframe-level solves each keyframe's rotation matrix. The feature-level uses pixel coordinates and inverse depth to obtain feature spatial coordinates. The observation-level is divided into two phases. The first phase uses coordinates from feature-level and the second phase uses rotation matrix from keyframe-level to calculate final Jacobian and residual. This three-level computation enables two unique SLAM data reuse. First, each keyframe's rotation matrix is reused over all observations within the keyframe. Second, each feature's coordinate is reused across its associated observations. Since the number of features is 10x more than keyframes, we prioritize feature reuse over keyframe reuse, thus calculating Jacobian matrix in feature (row)-stationary dataflow. Fig. 2c shows the circuit for IMU Jacobian, which consists of two pipeline stages. The first stage contains three parallel blocks for Jacobian matrix calculation, and the second stage calculates the residual and stores Jacobian and residual. Zero and identities of IMU Jacobian matrix will not be stored, which can reduce memory by 72%.

SLAM requires us to solve the linear system $\mathbf{A}\Delta\mathbf{p}=\mathbf{b}$. We use Schur elimination to simplify the equation, where the visual Jacobian matrix is divided into four blocks (Fig. 3a). Blocks **U**, **W**, and **X** only relate to visual observations, and **V** relates to IMU and prior information. Thus, when calculating Schur complement matrix $\mathbf{V} - \mathbf{W}\mathbf{U}^{-1}\mathbf{X}$, it can be considered that we first calculate the visual part and then add IMU and prior information to it. Two optimization schemes are proposed in the Schur elimination block. First, we make **U** as a diagonal matrix to reduce the computational complexity of $\mathbf{U}^{-1}$ from $O(n^3)$ to $O(n)$. Second, when **U** is a diagonal matrix, **X** becomes the transpose of **W**, reducing the on-chip memory storage requirement by 1.34x. After Schur elimination, Cholesky decomposition decomposes the symmetric matrix **S** into a lower triangular matrix **L** such that $\mathbf{L}\mathbf{L}^T=\mathbf{S}$. Fig. 3b illustrates the circuit for Cholesky Decomposition, where the hardware iteratively generates the *i*-th column of matrix **L** (Evaluate) and updates **S** for calculating (*i-1*)-th column of **L** (Update). We find that at *i*-th iteration, the number of operations of Evaluate and Update are *i* and *i*(*i-1*)/2, respectively. Thus, we propose to pipeline Evaluate and Update, where multiple Update units are time-multiplexed with the Evaluate unit. With pipelining and time-multiplexing, the latency is reduced by 5.75x with 3.3x less resources consumption.

Marginalization uses NLS solver outputs and performs $\mathbf{A} - \mathbf{Z}\mathbf{M}^{-1}\mathbf{Z}^T$ to generate the priors for the next window computation (Fig. 4a). The difficulty lies in $\mathbf{M}^{-1}$ computation. We propose to divide **M** into four blocks and make $\mathbf{M}_{11}$ as a diagonal matrix. In this way, the Schur elimination and Cholesky decomposition circuits for NLS solver can be reused in marginalization, greatly reducing resource consumption without performance degradation. During marginalization, **S** matrix that stores the parameters for linear system, contributes 60% of total memory (Fig. 3a). We notice **S** is a symmetric matrix, so the memory can be reduced by half. To further reduce the storage, we leverage the unique SLAM data structured sparsity. Since **S** is obtained by integrating camera and IMU, we propose to store their contributions separately. IMU's observation only relates to adjacent keyframes, so the non-zero elements are in diagonal and sub-diagonal blocks. The non-zero elements of camera contributions only exist in the 6x6 sub-block of each state, donating 6 DoF. The camera storage is further reduced by limiting the number of keyframes that capture the feature (co-observations). The storage is reduced by 4.1x in this process.

The design is dynamically optimized at runtime to adapt to different surroundings and save power while maintaining accuracy (Fig. 4b). When entering new environments with various feature points, the number of NLS iterations is dynamically adjusted to meet target accuracy based on the offline constructed lookup table. Along with NLS solver iterations, the number of Schur elimination modules and update modules during Cholesky decomposition will be dynamically reconfigured for less resource consumption. Since the lookup table is updated asynchronously, this runtime reconfiguration has minimal overhead. Instead of reconfiguring bitstream to FPGA, we applied clock gating for dynamically adjusted modules, enabling 1.59x power reduction with only 0.15% overhead. This runtime optimization has little impact on accuracy with <0.01cm degradation and sometimes even improves the accuracy due to its stochastic nature.

The proposed hardware is implemented on Xilinx ZC706 FPGA, with a fixed operational frequency at 143 MHz (Fig. 5a). We evaluate the design with two datasets: EuRoC for drones and KITTI Odometry for cars (Fig. 5b). Compared with CPU operating at 2.9 GHz, our FPGA design achieves 8.73x (10.49x) speedup and 164x (183x) energy reduction on EuRoC (KITTI). Compared with TX1 operating at 1.9 GHz, our FPGA design achieves 70x (45x) speedup and 41x (25x) energy reduction on EuRoC (KITTI). To validate the generalization of our design, we evaluate two additional Xilinx FPGAs: Kintex-7 and Virtix-7 series. Evaluated on EuRoC, our design achieves 7x and 11x speed up as well as 56x and 86x energy reduction over CPU on two boards. The significant efficiency gains are consistently found on KITTI dataset. Fig.6 demonstrates that our design achieves >5x better performance against recent prior SLAM accelerators.

**Acknowledgements:**

This work was supported in part by National Natural Science Foundation of China under Grant U21B2031, and C-BRIC, one of six centers in JUMP, a SRC program sponsored by DARPA.

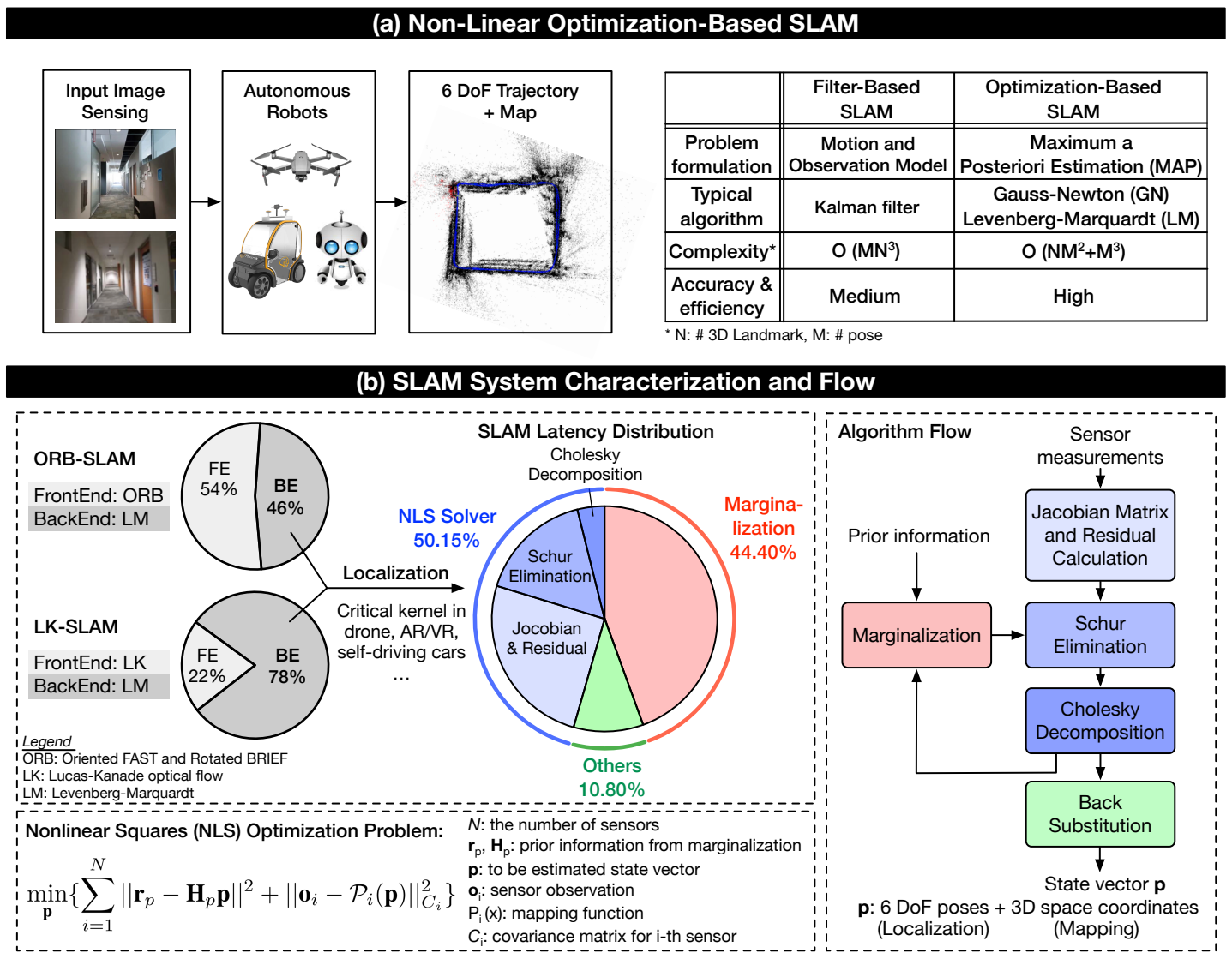

| | Filter-Based SLAM | Optimization-Based SLAM |
|---|---|---|
| Problem formulation | Motion and Observation Model | Maximum a Posteriori Estimation (MAP) |
| Typical algorithm | Kalman filter | Gauss-Newton (GN) Levenberg-Marquardt (LM) |
| Complexity* | O ($MN^3$) | O ($NM^2+M^3$) |
| Accuracy & efficiency | Medium | High |

* N: # 3D Landmark, M: # pose

Fig. 1. Robotic localization algorithm comparison, system profiling, and associated processing procedures.

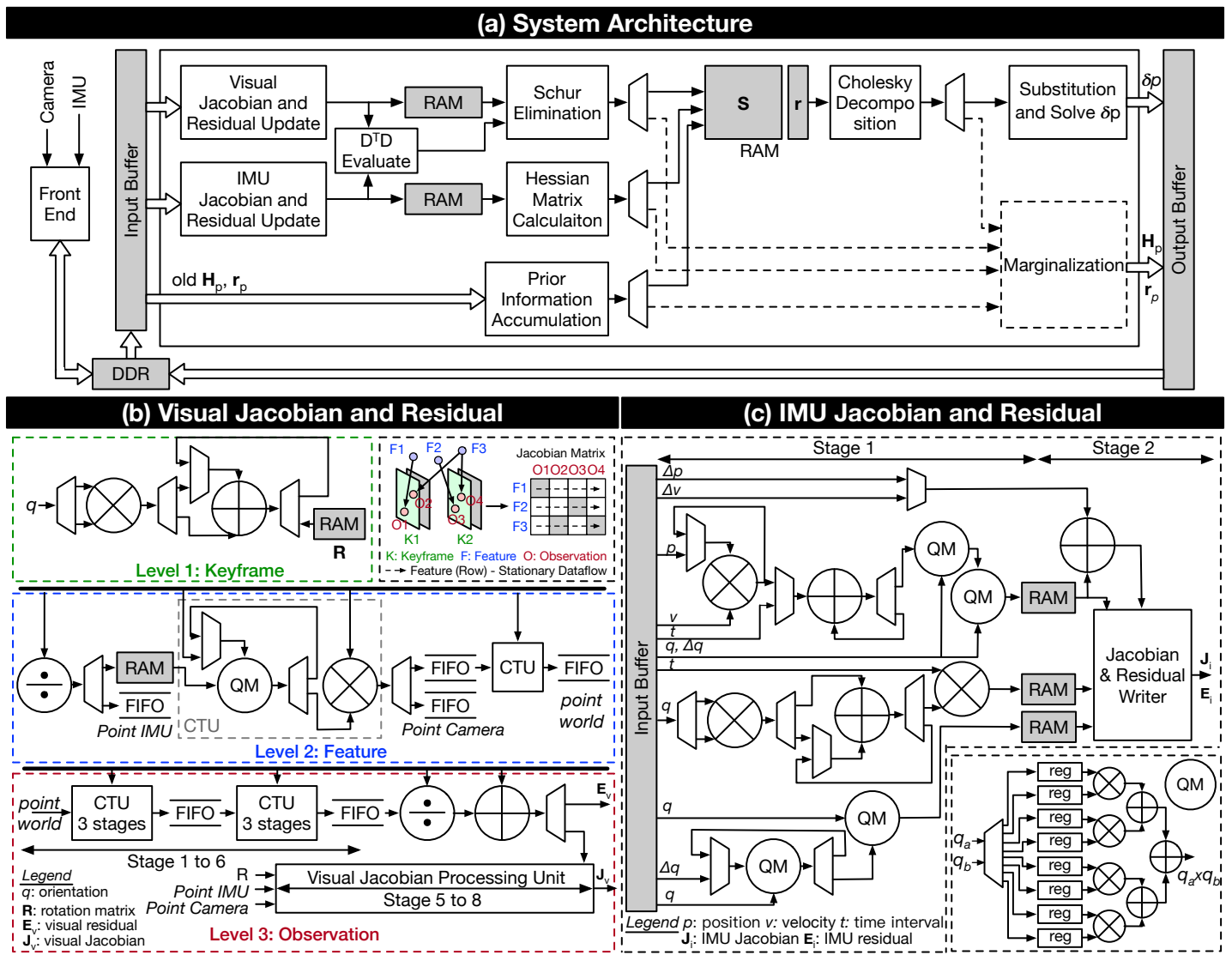

Fig. 2. Proposed overall robotic localization system architecture, with detailed Jacobian and Residual block for both vision and IMU.

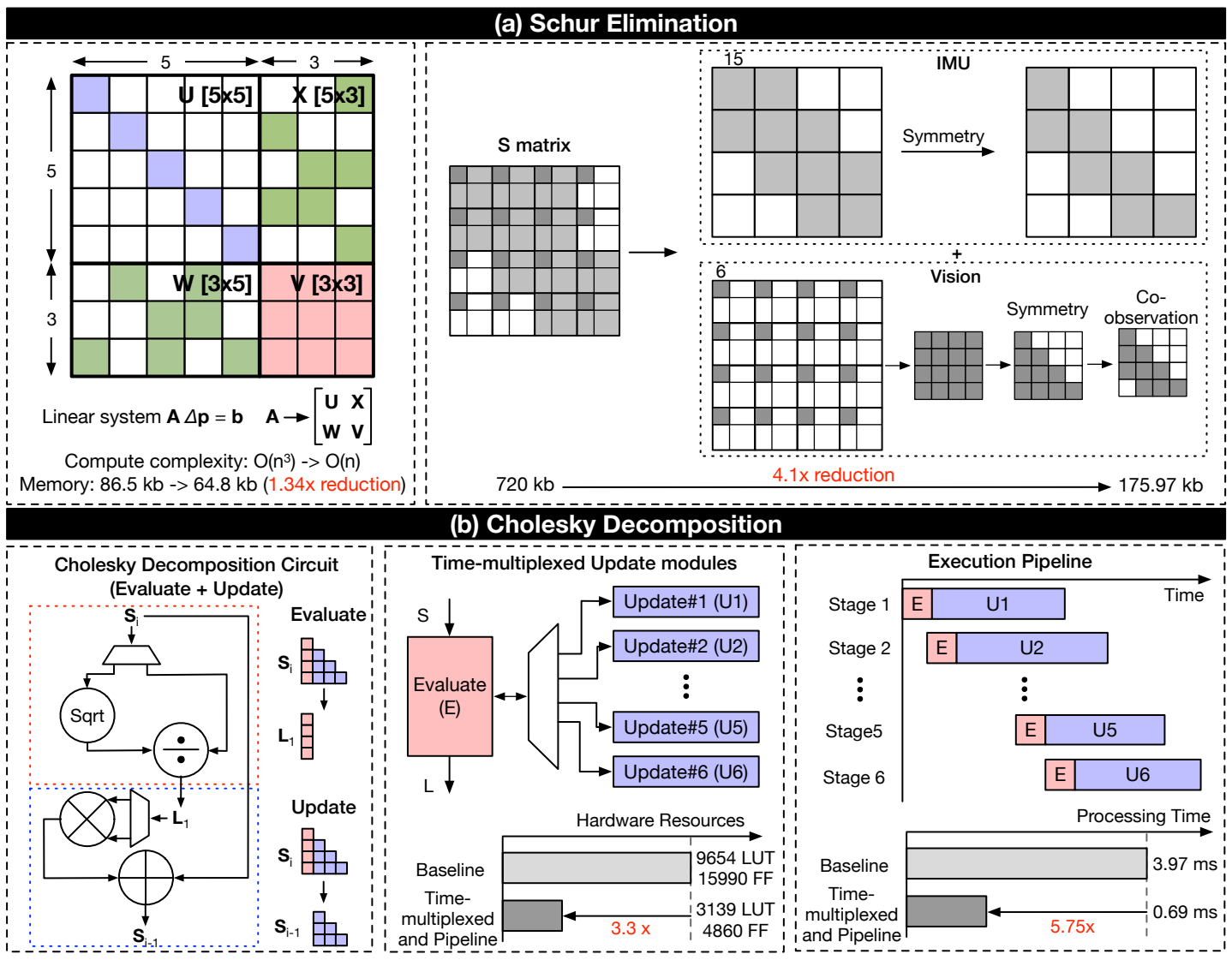

Fig. 3. Proposed memory optimization for Schur elimination, as well as time-multiplexed and pipeline for Cholesky decomposition.

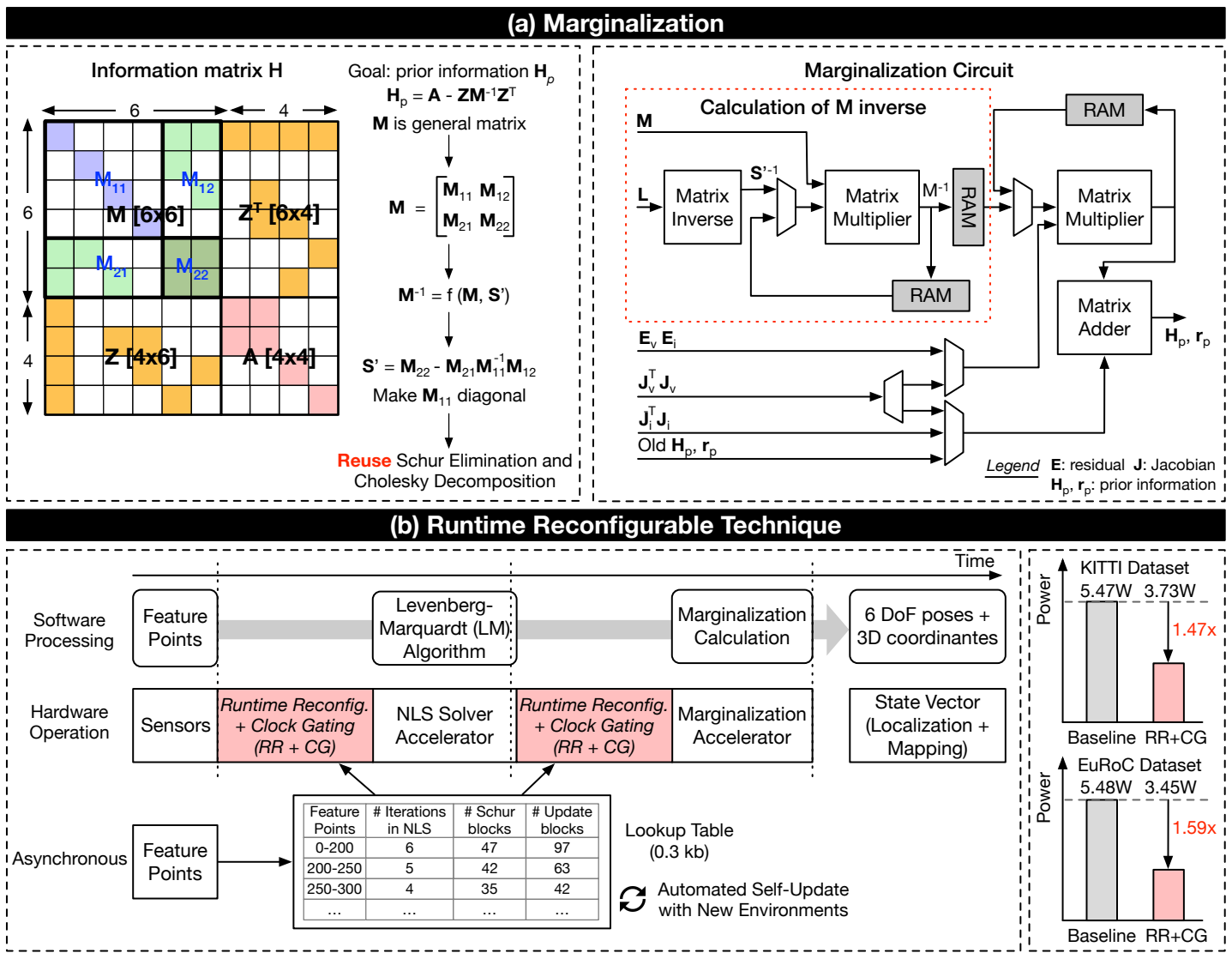

Fig. 4. Proposed hardware optimization for marginalization, and dynamic optimization techniques for robotic adaptive computing.

(a) FPGA Flatforms and Resource Utilization

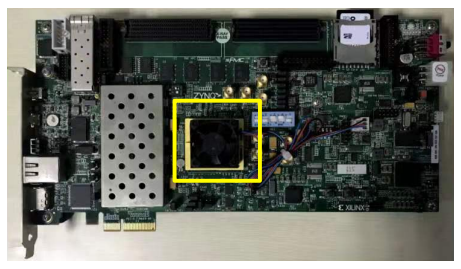

FPGA Zynq-7000 SoC ZC706 with XC7Z045 FFG900-2

| | LUT | FF | BRAM | DSP |
|---|---|---|---|---|
| FPGA ZC706 | 144108 (65.92%) | 172935 (39.56%) | 268 (49.17%) | 869 (96.56%) |
| Kintex-7 Series (XC7K160tfbg484) | 93670 (92.38%) | 123225 (60.76%) | 135 (41.38%) | 437 (72.83%) |
| Virtix-7 Series (XC7VX690tffg1761) | 232088 (53.58%) | 276284 (31.89%) | 425 (28.91%) | 1254 (34.83%) |

Kintex-7 Series is optimized for best price-performance.
Vertix-7 Series is optimized for highest system performance.

(b) Processing Latency and Energy of FPGA, CPU and GPU

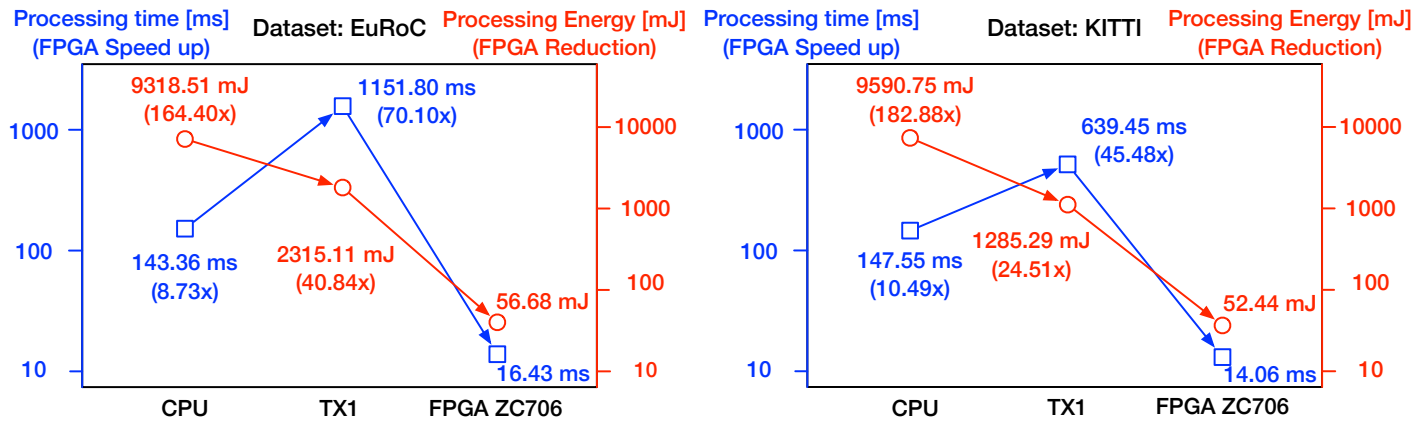

| EuRoC Dataset (For drone) | FPGA Speedup | | FPGA Energy Reduction | |
|---|---|---|---|---|
| | Over CPU | Over TX1 | Over CPU | Over TX1 |
| FPGA ZC706 | 8.73x | 70.10x | 164.40x | 40.84x |
| Kintex-7 Series (XC7K160tfbg484) | 7.01x | 56.30x | 180.73x | 44.90x |
| Virtix-7 Series (XC7VX690tffg1761) | 10.75x | 86.34x | 172.05x | 42.75x |

| KITTI Dataset (For car) | FPGA Speedup | | FPGA Energy Reduction | |
|---|---|---|---|---|
| | Over CPU | Over TX1 | Over CPU | Over TX1 |
| FPGA ZC706 | 10.49x | 45.48x | 182.88x | 24.51x |
| Kintex-7 Series (XC7K160tfbg484) | 8.27x | 35.82x | 196.09x | 26.28x |
| Virtix-7 Series (XC7VX690tffg1761) | 12.71x | 55.08x | 188.60x | 25.28x |

Fig. 5. Measurement on three FPGA platforms with two datasets, and performance and power comparison with CPU and TX1.

| | This work | ISSCC'19 CNN-SLAM [1] | JSSC'19 Navion [2] | TC'20 pi-BA [3] | RSS'17 VIO on Chip [4] | HPCA'21 Eudoxus [5] |
|---|---|---|---|---|---|---|
| Platform | FPGA | ASIC | ASIC | FPGA | FPGA | FPGA |
| Technology | 28 nm | 28 nm | 65 nm | 28nm | 28nm | 16nm |
| Design | digital | digital | digital | digital | digital | digital |
| Type | SLAM | SLAM | SLAM | SLAM | SLAM | SLAM |
| Algorithm | Levenberg-Marquardt (optimization-based) | Levenberg-Marquardt (optimization-based) | Gaussian-Newton (optimization-based) | Levenberg-Marquardt (optimization-based) | Gaussian-Newton (optimization-based) | Kalman Filter (Filter-based) |
| DoF | 6-DoF | 6-DoF | 6-DoF | 6-DoF | 6-DoF | 6-DoF |
| Voltage | 1 V | 0.63-0.9V | 1.2V | 1 V | 1 V | 0.85 V |
| Power | 3.45W | 243.6mW @ 0.9V 61.75mW @ 0.63V | 24mW | 5.50W | 1.46 W | 8.96W |
| Frequency | 143 MHz | 240 MHz | 62.5/83.3 MHz | 143 MHz | 100 MHz | 180 MHz |
| Throughput | 55.8 GOPS | 879.6 GOPS @ 0.9V 329.8 GOPS @ 0.63V | 10.5-59.1 GOPS | N/A | 4.4-24.6 GOPS | N/A |
| Latency | 16.43 ms | N/A | 30.8 ms | 110 ms | 200 ms | 44.6 ms |
| Energy per Frame | 56.6 mJ | N/A | 739.2 uJ | 605 mJ | 292 mJ | 399.6 mJ |
| Dynamic Optimization | Yes | N/A | N/A | No | No | No |

Fig. 6. Comparison with recent prior works.